\pdfoutput=1 
\documentclass[prl,twocolumn,a4paper,showpacs,aps,superscriptaddress]{revtex4}
\usepackage{amsmath,amsfonts,bm,graphicx,dsfont,overpic}
\usepackage{color}
\usepackage{overpic}

\newcommand{\hrho}{{\rho}}
\newcommand{\leads}{l}
\newcommand{\ext}{ext}
\newcommand{\Tr}{\mathsf{Tr}}

\newcommand{\tot}{\mathsf{tot}}

\newcommand{\gate}{\mathsf{g}}

\newcommand{\Tab}[1]{\mbox{Tab.\ \ref{#1}}}

\newcommand{\Fig}[1]{\mbox{Fig.\ \ref{#1}}}

\begin{document}

\title{Sources of negative tunneling magneto-resistance in multilevel quantum dots with ferromagnetic contacts}
\author{Sonja Koller}
\affiliation{Theoretische Physik, Universit\"at Regensburg, 93040 Regensburg, Germany}
\author{Jens Paaske}
\affiliation{The Niels Bohr Institute and Nano-Science Center, University of Copenhagen, Universitetsparken 5, DK-2100 Copenhagen {\O}, Denmark}
\author{Milena Grifoni}
\affiliation{Theoretische Physik, Universit\"at Regensburg, 93040 Regensburg, Germany}

\date{\today}

 \begin{abstract}
We analyze distinct sources  of spin-dependent energy level shifts and their impact on the tunneling magnetoresistance (TMR) of interacting quantum dots coupled to collinearly polarized ferromagnetic leads. Level shifts  due to virtual charge fluctuations can be quantitatively evaluated within a diagrammatic representation of our transport theory.
The theory is valid for multilevel quantum dot systems and we exemplarily apply it to carbon nanotube quantum dots, where we show that the presence of many levels can qualitatively influence the TMR effect.
 \end{abstract}


\maketitle
Recent transport experiments on quantum dots coupled to ferromagnetic leads have demonstrated the existence of spin-dependent  energy level shifts, through the observation of negative tunnel magneto-resistance (TMR) effects in the single electron tunneling regime~\cite{Sahoo05},
 and spin splitting in the Kondo regime~\cite{Pasupathy04,Hauptmann08,Hofstetter10,Gaas11}.
So far qualitative different approaches for explaining the origin of the underlying shifts co-exist. For example, negative TMR data from CNT measurements have been fitted with a model relying on spin dependent interfacial phase shifts ~\cite{Cottet06b} picked up by the wave function during multiple reflections at a spin-active interface \cite{Cottet06}.
  The concept is related to that of spin-mixing conductance~\cite{Brataas99} and, because it  only depends on the properties of  the spin-active barrier region,    is only weakly gate dependent and present both in interacting and in non-interacting systems~\cite{Wetzels05}.
  In contrast, experiments on CNTs ~\cite{Hauptmann08,Gaas11} and InAs nanowires \cite{Hofstetter10} in the Kondo regime  have demonstrated a combination of gate dependent and gate independent contributions to the energy level shifts. Both can be explained in terms of charge fluctuations \cite{Haldane78}, whereby  electron-electron interactions are responsible for the logarithmic gate dependence  \cite{Martinek03}, while a Stoner splitting of the energy bands of the magnetically polarized leads
  accounts for the almost gate independent part \cite{Martinek05}.

 While the effects of the energy level shifts in the Kondo regime are by now well understood, a thorough understanding of their  influence on the TMR phenomenon in interacting quantum dots is still missing. The negative TMR data  \cite{Sahoo05} have been satisfactorily fitted in terms of a generalized Anderson model already including   gate-independent level shifts.
 In  \cite{Koller07} a reflection Hamiltonian was included to account for reflection processes at the interface in a second order sequential tunneling theory.

 In this work, we specifically address the TMR phenomenon  and discuss different level-shift induced mechanisms yielding negative TMR.
In particular we discuss  two types of spin-dependent level shifts that arise intrinsically from tunneling induced renormalization: strongly gate-dependent ones from a net difference of majority and minority density of states at the Fermi level,   and largely
 gate-independent ones from a Stoner shift of the majority and minority bands in the leads. We show how to include these effects within a diagrammatic approach to the reduced density matrix of the nanosystem \cite{Schoeller94} for
 generic multilevel quantum dots, and reproduce the  results of \cite{Martinek05} in the case of a simple Anderson model.
 Finally,
we analyze
the TMR of a CNT quantum dot, and show that, due to the multilevel spectrum, intrinsic contributions only  can yield
a marked gate voltage dependence and a TMR which can indeed become negative.

\begin{figure}\begin{center}
\includegraphics[width=0.9\columnwidth]{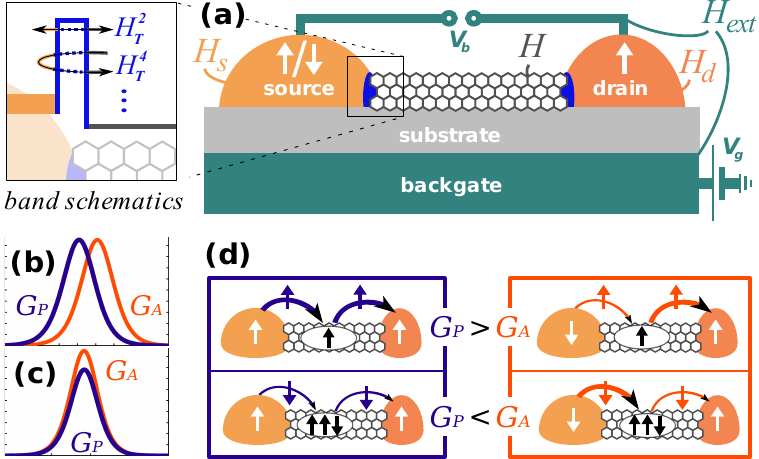}
\caption{\label{fig:setup}(a) Transport set-up of a carbon nanotube (CNT) quantum dot with ferromagnetic leads. If an electron enters the barrier at the tube ends, it can either tunnel to the lead or being reflected at the interface. Tunneling processes in the contact region are described by even powers of the tunneling Hamiltonian $H_T$. (b)/(c) Conductances $G_P, G_A$ versus gate voltage for two situations exhibiting negative TMR. (d) The negative TMR mechanism related to (c), where thick/thin lines describe processes that are favoured/disfavoured being associated to majority/minority spins. }\vspace{-1cm}
\end{center}\end{figure}

\paragraph{Mechanisms of negative TMR.}
The basic mechanism underlying a negative TMR is the presence of an effective generalized Zeeman
field  $h^{P/A}_{}(V_g)$ accounting for both extrinsic (stray fields, applied magnetic field)
and intrinsic 
sources of spin-dependent level shifts. Here ($P/A$) refers to contacts with parallel/antiparall magnetization, see Fig. \ref{fig:setup}.
 Intuitively, one would expect that the  conductance for contacts polarized in parallel ($G_{P}$) is larger than the one for the anti-parallel case ($G_{A}$), yielding positive values of $\mathrm{TMR}:=(G_{P}/G_{A})-1$. Nevertheless, there are at least two  different mechanisms which can lead to a negative TMR, as
 %
sketched in Figs. \ref{fig:setup}(b), (c), respectively. Here
 we consider quantum dots in the single electron tunneling regime and look at the linear conductance  as a function of gate voltage $V_g$.  For sufficiently small couplings the temperature determines the width of the conductance peaks, while the peak position signals at which value of  $V_g$ the two many-body  ground state configurations with  $N$ and $N+1$ electrons involved in the transition have the same electrochemical potential: $\mu(N,V_g)=\mu (N+1,V_g)$.
 Because the effective Zeeman field leads to corrections to the energy difference $E_b-E_a:=E_{ba}$ associated to the transition from a many-body state $a$ to a many-body state $b$, this in turn modifies  the position of the conductance peaks.

First, and most obviously, there can be negative TMR if there is a noticeable offset in the conductance peak positions in parallel and anti-parallel configuration, see \Fig{fig:setup}(b). This requires $|h^{P} - h^{A} |$ of the order of the width $k_BT$ of the conductance peaks. Secondly, however, negative TMR can also arise for 
$|h^{P}|=|h^A|\approx k_BT$, 
see \Fig{fig:setup}(c). This is because the effective magnetic field 
responsible for the effective Zeeman shift also removes the spin-degeneracy of the ground states  by favoring the states with maximum total spin.  
This situation is illustrated in Fig. \ref{fig:setup}(d) for the case of a CNT quantum dot, for which we consider the CNT  Hamiltonian Eq. (\ref{eq:armchair}).  The four-electron shells of the dot are filled sequentially upon sweeping the gate voltage and only a single shell needs  to be considered at a time. The case of orbital ($\Delta \varepsilon=0$) and spin degeneracy is illustrated for the $n$-th shell in Table I. For the $4n\leftrightarrow 4n+1$ and $4n+1 \leftrightarrow 4n+2$ transitions and $h^{P/A}\neq 0$  only spin-up electrons are required. Hence, due to the larger density of spin up electrons,  peak heights for parallel contact polarization will exceed those for the anti-parallel case. On the other hand, for the
  $4n+2\leftrightarrow 4n+3$ and $4n+3\leftrightarrow 4n+4$ transitions  spin-down electrons have to be transferred through the dot. In this latter case a configuration with anti-parallel contact polarization might be favored leading to $G_A>G_P$.
%
%
\paragraph{The model Hamiltonian} For a quantitative description we  consider the transport setup of \Fig{fig:setup}.
In the limit of weak coupling to the leads it can be described by the total Hamiltonian
\begin{equation}
\hat H^{P/A}_\tot=\hat H+\hat H^{P/A}_{\ext}+\sum_{l=s,d} \hat H_{\leads}+\hat H^{}_T,
\label{eq:ham}
\end{equation} where  $\hat H=\hat H_0 -e\alpha V_g \hat N$ comprises the Hamiltonian $\hat H_0$ of the isolated quantum dot
and the effects of a gate voltage ($\alpha$ is a conversion factor of the order of one).
In the case of an armchair nanotube quantum dot of medium-to-large radius far from half-filling it reads \cite{Mayrhofer06}
\begin{equation}
\hat H_0=\frac{1}{2}E_c \hat N^2 + \frac{1}{2}\sum_{\tilde r,\sigma}(\varepsilon_0 \hat N_{\tilde r \sigma} +r\Delta\varepsilon
)\hat N_{\tilde r \sigma} +\hat H_B, \label{eq:armchair}
\end{equation}
where $ \varepsilon_0=\hbar v_{\rm F}/\pi L$, with $v_{\rm F}$ the Fermi velocity
 and $L$ the CNT length, is the level spacing; $\Delta\varepsilon$ is the orbital mismatch, $E_c$ is the charging energy and $\hat H_B$ accounts for bosonic charge and spin excitations.
 The number of electrons in orbital band $\tilde r$ with spin $\sigma$ is $N_{\tilde r \sigma}$ and the total number is $N=\sum_{\tilde{r}\sigma}N_{\tilde{r}\sigma}$. The groundstates $\mid N_{+\uparrow}, N_{+\downarrow},N_{-\uparrow}, N_{-\uparrow}\rangle$ of shell $n$ have $4n$, $4n+1$, $4n+2$ and $4n+3$ electrons and
    can be characterized in terms of the excess spins in band $\{\tilde r,\sigma\}$ with respect to the case of equally filled bands:
    $\mid n, n , n, n \rangle:=\mid \cdot,\cdot \rangle$, $\mid n+1, n,n, n\rangle := \mid \uparrow,\cdot\rangle$, or $\mid n+1, n+1, n, n \rangle:=\mid \uparrow \downarrow,\cdot\rangle$, etc., see Table 1.
For medium-to-large tube radius and far from the charge neutrality point spin-orbit coupling and exchange effects are not relevant. They can become of interest for large curvatures and will be investigated elsewhere.
 $\hat H^{P/A}_{\ext}$ accounts for external, gate independent sources of level splitting.

 The leads are described by
$\hat H_{\leads}=\sum_{\sigma q}\epsilon_{l q}\hat c^\dag_{l\sigma q}\hat c_{l\sigma q}$, with $\hat c_{l\sigma q}$ annihilating an electron of energy $\epsilon_{l q}$ and of spin $\sigma$ in lead $l$.
We assume a constant density of states in lead $l$,  $D^{}_{l\sigma}$, which determines the leads  polarizations $P_l$ at the Fermi level according to  $P_l=(D_{l\uparrow}-D_{l\downarrow})/(D_{l\uparrow}+D_{l\downarrow})$.
%
On the other hand, in order to account for a Stoner splitting $\Delta_{\rm St}$, the range of available energies in the leads is given by \cite{Martinek05}
$-(W+\sigma \Delta_{\rm St}/2)\le \omega \le W-\sigma\Delta_{\rm St}/2$, where $W$ is the bandwidth at zero Stoner splitting.
Finally the perturbative contribution is $\hat H^{}_T=\sum_{l\sigma}\int\!\! d^3 r\left(T_{l}(\vec{r})\hat\psi^\dag_{\sigma}(\vec{r})\hat\phi^{}_{l\sigma}(\vec{r})+h.c.\right)$, allowing for tunneling between CNT and leads. Here $T_{l}(\vec{r})$  is the  tunnel coupling,  $\hat\psi_{\sigma}(\vec{r})$ the CNT bulk electron operator as given in Ref.~\cite{Mayrhofer06}, and $\hat\phi^{}_{l\sigma}(\vec{r})=\int d\epsilon\, D^{}_{l\sigma}(\epsilon)\sum_{q|_\epsilon}\phi_{lq}(\vec{r})\hat c_{l\sigma q}$ the lead electron operator with $\phi_{lq}(\vec{r})$ denoting the corresponding single-particle wave function.
%
\paragraph{Transport theory.}
The current, as any other observable of the transport set-up, can be calculated by a trace over the associated operator multiplied by the reduced density matrix $\hat\hrho (t)= \Tr_{\leads}\left\{\hat\hrho_\tot(t)\right\}$. Being obtained from the density matrix $\hat\hrho_\tot$ of the total set-up upon tracing out the lead degrees of freedom, $\hat\hrho(t)$ stores the full information about the state of the dot in the presence of the tunnel coupling to the leads. The time evolution of $\hat\hrho(t)$ follows the Liouville equation, and it can for the stationary state ($\dot{\hat\hrho}(t)=0$) be cast into the form, see e.g.~\cite{Koller10},
\begin{equation}
  0 = -i\sum_{aa'} \delta_{ab}\,\delta_{a'b'}\left(E_a-E_{a'}\right)
  \rho_{aa'}+\sum_{aa'}{{K}}^{aa'}_{bb'}\rho_{aa'},
  \label{vectoreq}
\end{equation}
 taking matrix elements with respect to the many-body eigenstates of the CNT: $\rho_{aa'}:=\langle a | \hat\rho | a'\rangle$ and $E_a:=\langle a |\hat H|a\rangle$. Furthermore $K^{aa'}_{bb'}:=\langle b | [K\langle a \rangle\langle a'|]| b'\rangle$, with the kernel superoperator $K$ arising from the perturbation, i.e., the tunnel coupling between quantum dot and leads. Depending on up to which  order $2n$ in the tunnel coupling $K$ is calculated, one takes into account effects from $2n$ correlated  tunneling events.
The most involved part of a perturbative transport calculation is the determination of the kernel $K$.
 In the supplemental material  we identify specific terms in all orders of the perturbation series as \emph{charge fluctuation processes}, which sum up to a Taylor series yielding the \emph{intrinsic} level renormalization $E_{ab}\to E_{ab}+h_{int}^{ab}$ to the energy difference $E_{ab}$, where
\begin{multline}h_{int}^{ba}=\sum_{l}\left(\sum_{c}\Bigl|T_{l\sigma}^{p}(c,b)\Bigr|^2
\int' d\omega\frac{f(\omega)D_{l\sigma}(\omega)}{\omega-\beta E_{cb}-p\beta e V_l}\right.\\
\left.
- \sum_{c'}\Bigl|T_{l\sigma}^{p}(c',a)\Bigr|^2
\int' d\omega\frac{(1-f(\omega))D_{l\sigma}(\omega)}{\omega-\beta E_{ac}+p\beta e V_l}\right).
\label{splitting}
\end{multline}
       We notice that
     Eq. (\ref{splitting}) generalizes to  complex quantum dots the  analysis results for the energy level shifts obtained for the single impurity Anderson model \cite{Haldane78,Martinek03,Martinek05}.
     It is convenient to rewrite $h_{int}^{ba}=h_{0}^{ba}(V_\gate)+\Delta h^{ba}(\Delta_{\rm St})$, where
     $\Delta h_{}^{ba}(\Delta_{\rm St} =0)=0$.
      As shown in the supplemental material,
      the first term  depends logarithmically on the gate voltage and the interaction strength, while
      $\Delta h^{ba}$ depends on the finite bandwidth $W_\sigma=W-\sigma\Delta_{\rm St}/2$ of $D_{l\sigma}(\omega)$ and will be almost gate and interaction independent and thus contribute in the same way as contributions $h^{P/A}_{ext}$ induced by $\hat{H}^{P/A}_{ext}$.

      Notice that for quantum dots with PdNi contacts, given a tunneling coupling $\Gamma \simeq 1$ meV, one gets $\Delta h^{ba}$   of the order of few hundreds
$\mu$eV \cite{Gaas11}. This is the same order of magnitude as that of the
        effective constant field used
       to fit the TMR  data in Ref. \cite{Sahoo05} with the interface phase shift model \cite{Cottet06b}
        ( $\Gamma =\sum_l \Gamma_l $ and $\Gamma_l\sim D_l \mid T_l \mid^2$).
        A full account of the interface phase shifts might be necessary for large transparencies and goes beyond the scope of our tunneling Hamiltonian.
        However, in the regime where the tunnelling model is valid (Coulomb blockade and Kondo regime), tunneling induced shifts should suffice to explain the experiments.

\begin{figure*}\begin{center}\vskip 0.24cm
\begin{tabular}{@{\hspace{-1em}}c@{\hspace{-1em}}c@{\hspace{-1em}}c}
\includegraphics[width=0.72\columnwidth]{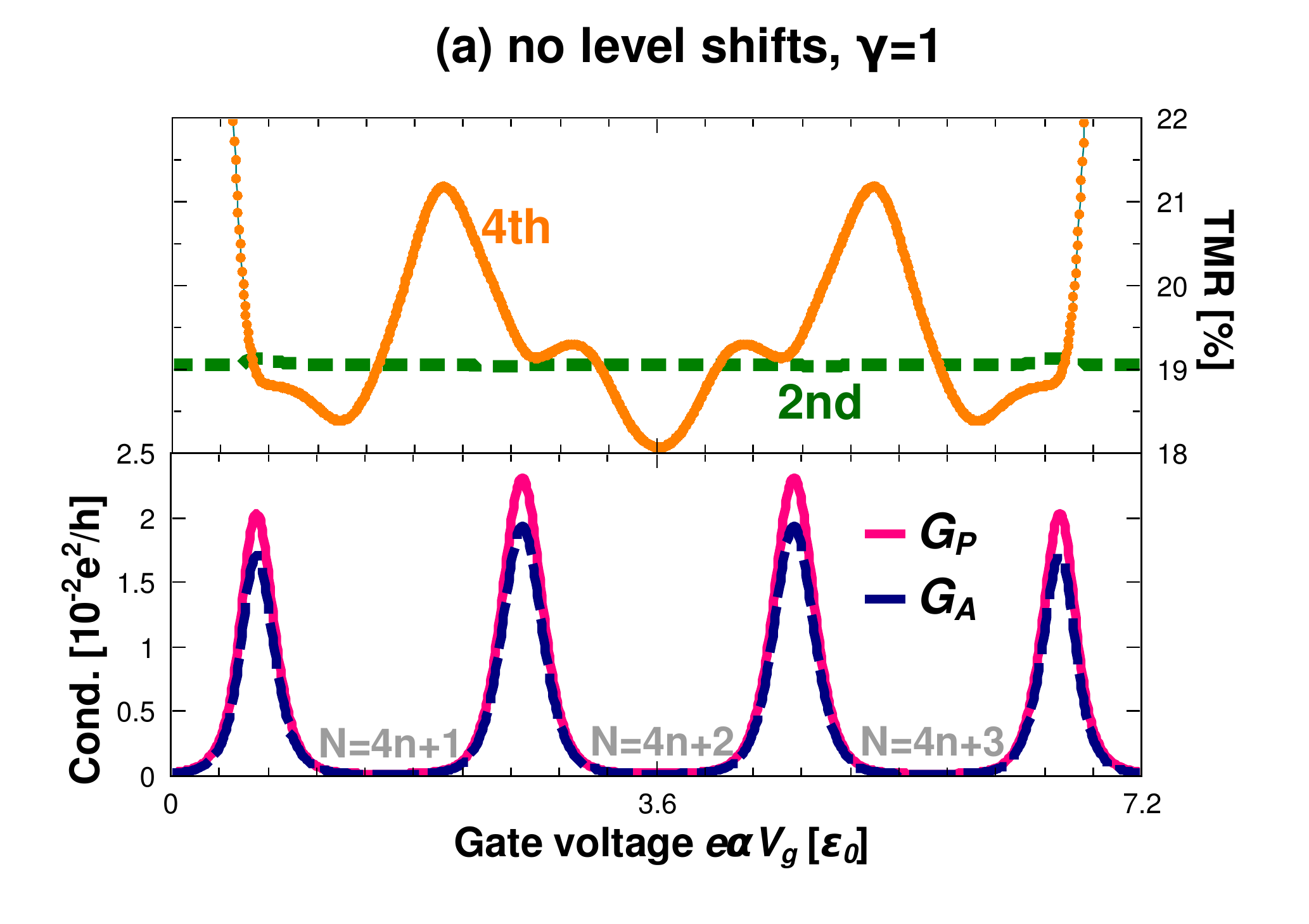}&\includegraphics[width=0.72\columnwidth]{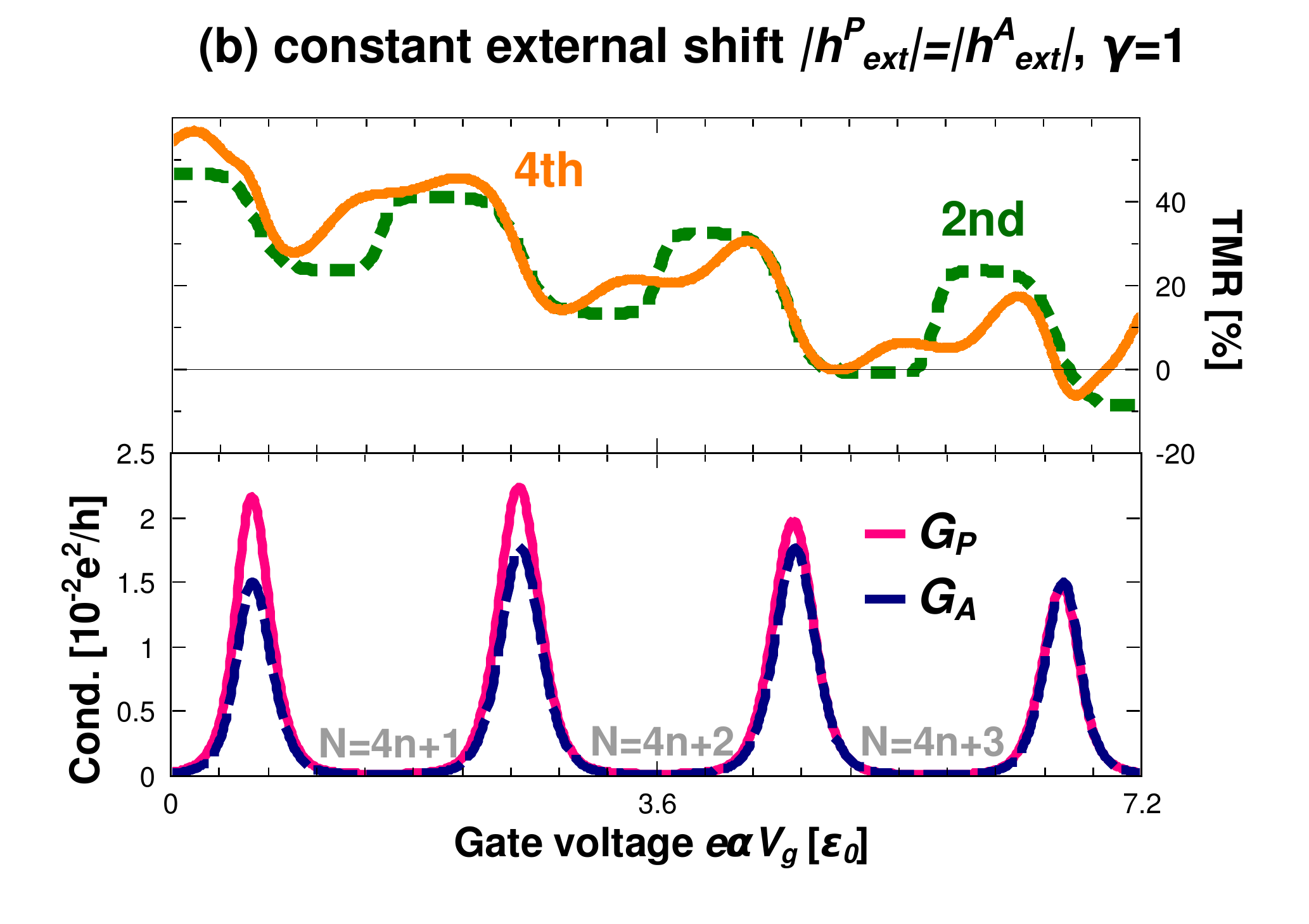}&\includegraphics[width=0.72\columnwidth]{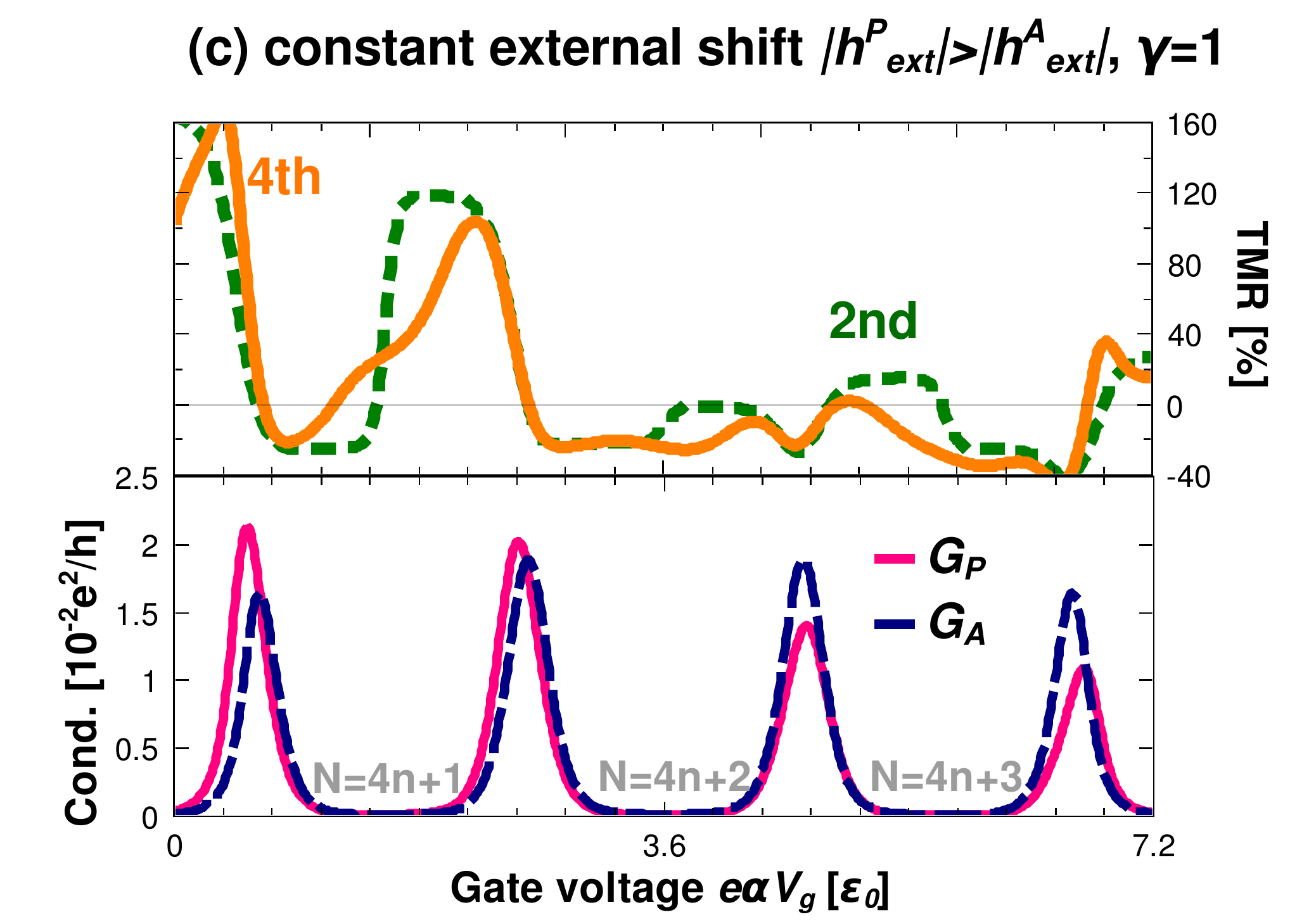}\\
\includegraphics[width=0.72\columnwidth]{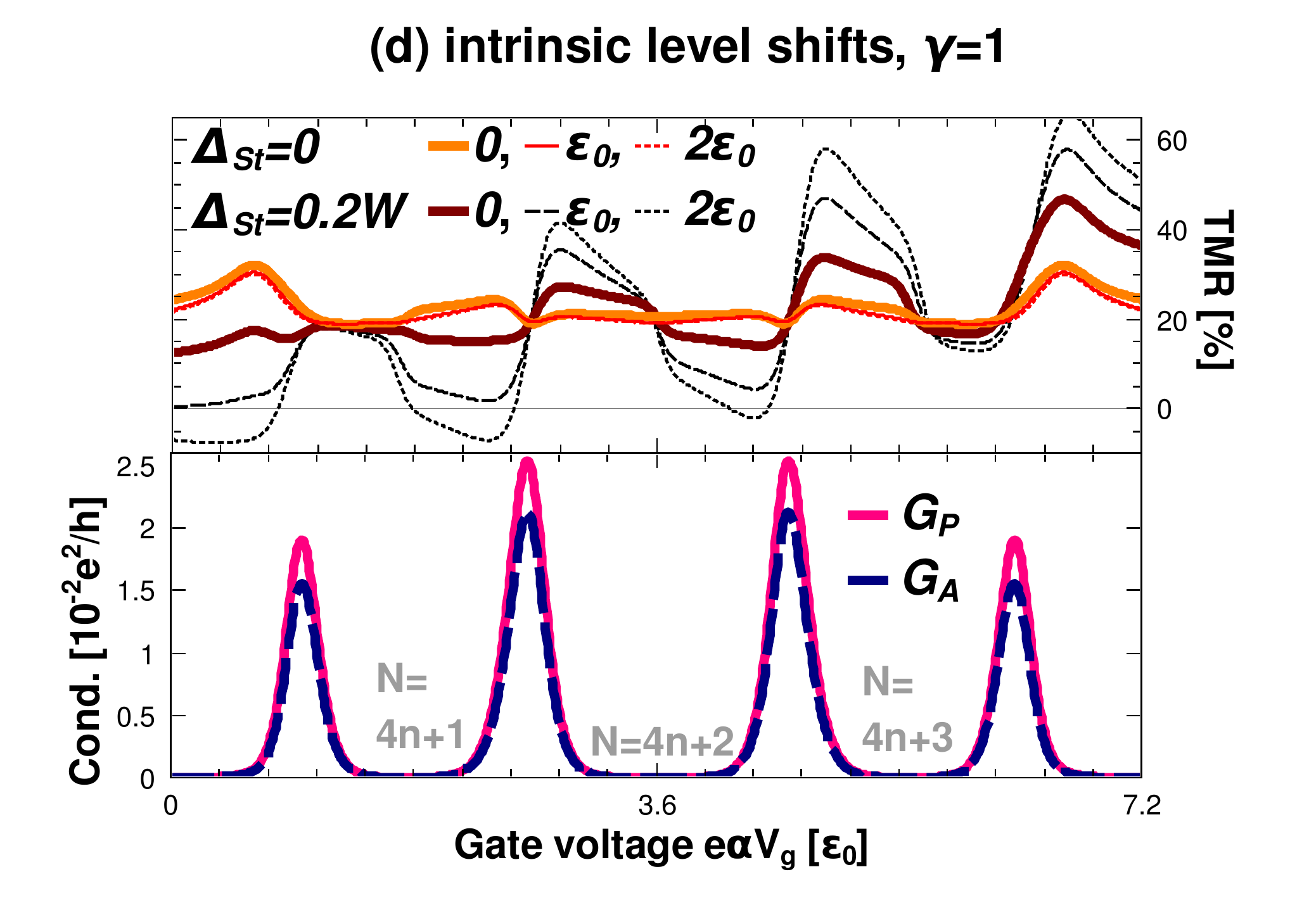}&\begin{overpic}[width=0.72\columnwidth]{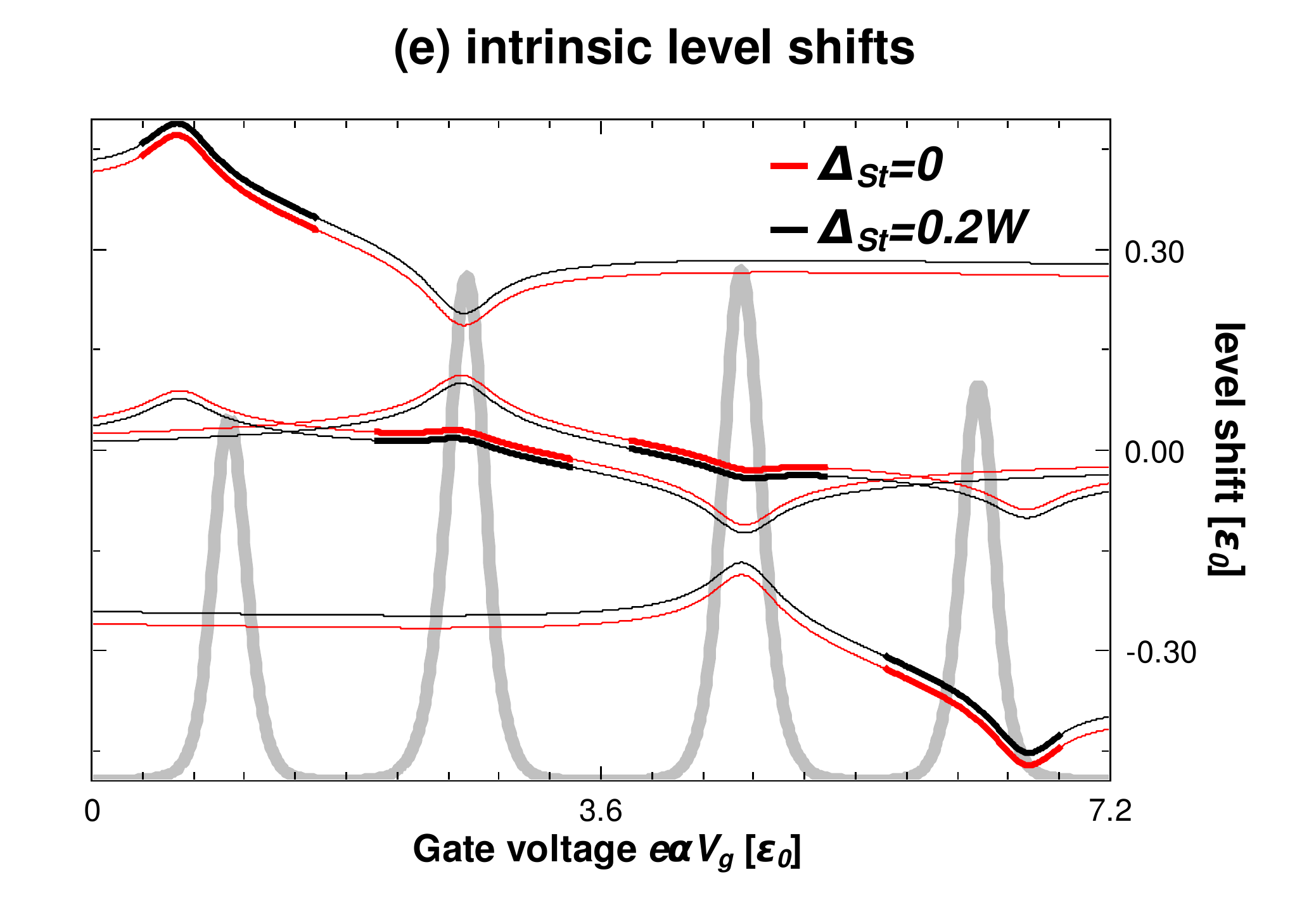}\put(9,54){\mbox{\tiny{$|\cdot,\cdot\rangle$}}}
\put(9,50){\mbox{\tiny{$\to|\uparrow,\cdot\rangle$}}}
\put(22,31){\mbox{\tiny{$|\uparrow,\cdot\rangle\to|\uparrow,\downarrow\rangle$}}}
\put(45,39){\mbox{\tiny{$|\uparrow,\downarrow\rangle\to|\uparrow\downarrow,\downarrow\rangle$}}}
\put(65,24){\mbox{\tiny{$|\uparrow\downarrow,\downarrow\rangle\to$}}}
\put(70,20){\mbox{\tiny{$|\uparrow\downarrow,\uparrow\downarrow\rangle$}}}\end{overpic}&\includegraphics[width=0.72\columnwidth]{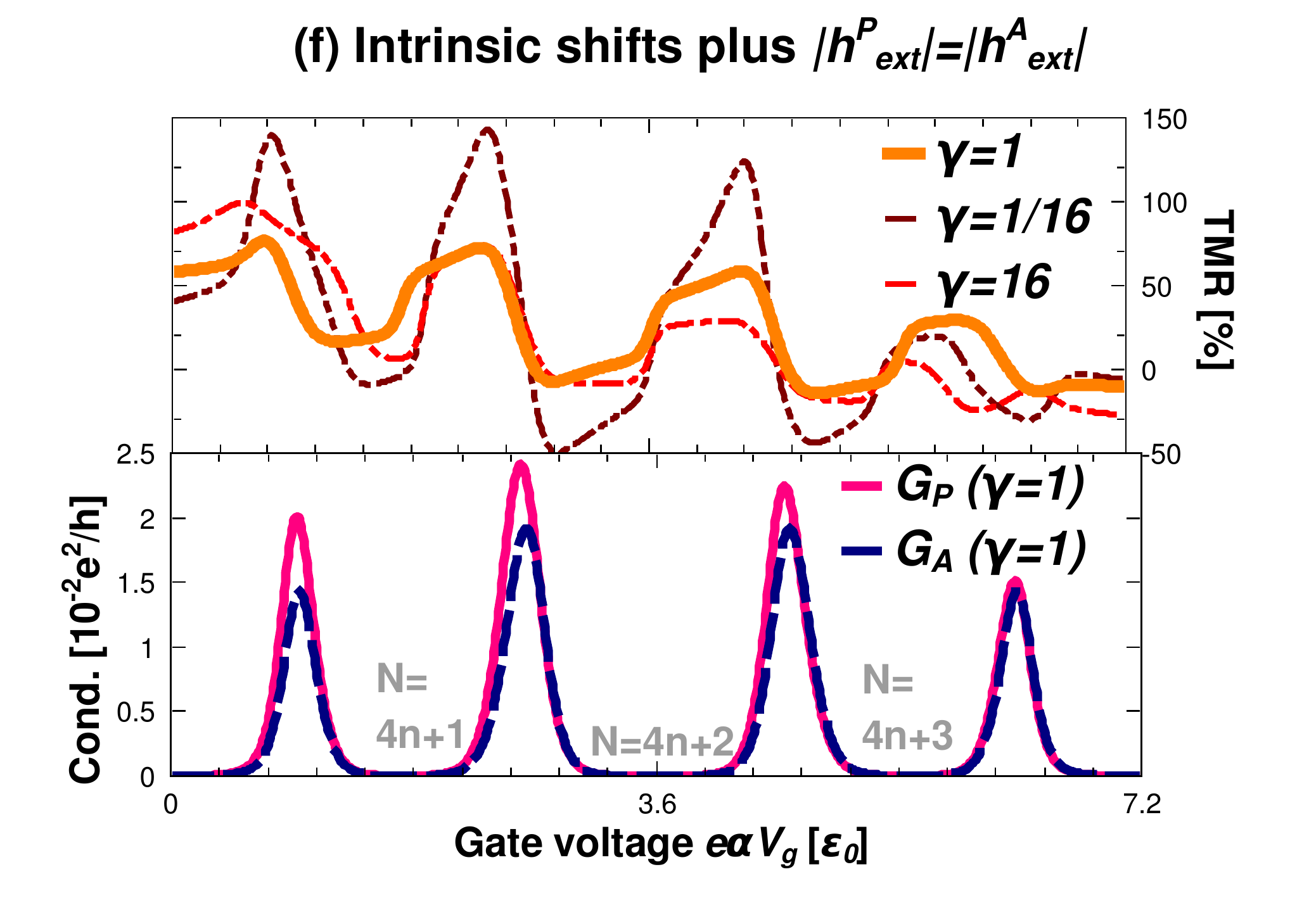}
\end{tabular}\vspace{-0.5cm}
\caption{\label{fig:TMRs} Parallel ($G_{P}$) and anti-parallel ($G_{A}$) conductance along with the resulting tunnelling magneto-resistance (TMR) for a CNT of $500\,$nm length ($\varepsilon_0=3.35\,$meV) and charging energy $E_c=6.7\,$meV. The thermal energy was set to $k_BT=0.3\,$meV, the lead polarization to $P=0.4$, the tunnel broadening to $\hbar\Gamma_s=3\,\mu$eV and the bandwidth to $W=3$eV. (a) For a full fourth order calculation a mirror-symmetric TMR slightly oscillating around a value of $20\%$ is obtained. (b) An equal splitting $h^P_{ext}=h^A_{ext}=0.4\,$meV causes gate asymmetry and negative values of the TMR by the mechanism \Fig{fig:setup}(c). (c) For $h^P_{ext}=2h^A_{ext}=0.8\,$meV, the regularity of the curve is broken and the negative TMR mechanism \Fig{fig:setup}(b) comes into play. (d) TMR for $\Delta_{\rm st}=0, 0.2W$ and conductance for $\Delta_{\rm St}=0$.   Including the influence of the excited states (with cut-off  set to $\epsilon_0$ and $2\epsilon_0)$ negative TMR can be reached.
With $\Delta_{\rm St}=0$ electron-hole-symmetry is preserved, for $\Delta_{\rm St}\neq0$ is broken.
(e) This  is confirmed by direct investigation of the intrinsic level shifts for some electron-hole symmetric transitions in parallel configuration. (f) Combining intrinsic shifts with external splitting. An asymmetric coupling to the leads can enlarge ($\gamma<1$) or diminish ($\gamma>1$) the TMR effect.
}
\end{center}\vspace{-0.5cm}\end{figure*}

\begin{table}
\begin{tabular}{c@{\hspace{1em}}|@{\hspace{1em}}c@{\hspace{1em}}}
filling & ground state(s)\\\hline\hline
$N=4n$ & \colorbox{yellow}{$|\cdot,\cdot\rangle$}   \\\hline
$N=4n+1$ & \colorbox{yellow}{$|\uparrow,\cdot\rangle,|\cdot,\uparrow\rangle$},$|\downarrow,\cdot\rangle,|\cdot,\downarrow\rangle$ \\\hline
$N=4n+2$ & \colorbox{yellow}{$|\uparrow,\uparrow\rangle$},$|\uparrow\downarrow,\cdot\rangle,|\uparrow,\downarrow\rangle,|\cdot,\uparrow\downarrow\rangle,|\downarrow,\uparrow\rangle,|\downarrow,\downarrow\rangle$ \\\hline
$N=4n+3$ & \colorbox{yellow}{$|\uparrow\downarrow,\uparrow\rangle,|\uparrow,\uparrow\downarrow\rangle$},$|\uparrow\downarrow,\downarrow\rangle,|\downarrow,\uparrow\downarrow\rangle$ \\\hline
\end{tabular}{\mbox{\tiny{$|\cdot,\cdot\rangle$}}}
\caption{Ground states of a  CNT in shell $n$ accounting for spin $\sigma= \{\uparrow, \downarrow\}$ and orbital degrees of freedom $\tilde{r}=\{+/-\}$, leading to four-electron shells.  The number of electrons in band $\tilde r$ with spin $\sigma$ is $N_{\tilde r \sigma}$ and in the dot $N=\sum_{\tilde{r}\sigma}N_{\tilde{r}\sigma}$.
  With $h^{P/A}\neq0$, only the highlighted states are ground states.
    }
\label{states}
\end{table}
\paragraph{Linear conductance results.} We consider the linear bias regime for spin polarized transport across a CNT without band-offset, cf. Table 1.
\Fig{fig:TMRs} shows the conductances  for parallel (pink solid) and anti-parallel (blue dashed) lead magnetizations under the influence of the distinct types of level shifts, along with the corresponding TMR curves.
Throughout this work we assume equal lead polarizations  $P_s=P_d=P$ and a coupling asymmetry $\gamma=\Gamma_s/\Gamma_d\sim |T_s|^2/|T_d|^2$, with the actual values of  $\gamma$ given at each plot.\\
In the \emph{absence} of any Zeeman shift, \Fig{fig:TMRs}(a) shows the typical conductance peak patterns expected for the degeneracies according to \Tab{states}, for both a second order (sequential tunneling) and a fourth order (cotunneling, pair tunneling, single charge fluctuations etc.) \cite{Koller10} truncation in the calculation of the kernel $K$. Second order  theory predicts  a constant TMR value~\cite{Koller07}, 
but this changes upon inclusion of higher order effects. To \emph{fourth} order in perturbation theory the TMR exhibits an oscillatory gate voltage dependence, albeit the variation is small \cite{Weymann05}.
To visualize the TMR mechanism from \Fig{fig:setup}(c) and (d), we include in \Fig{fig:TMRs}(b) merely a constant shift from an external source, $|h^{P}_{ext}|=|h^{A}_{ext}|>k_BT$ (this can arise from the external field which controls the direction of contact polarization in TMR measurements, when for the measurement of $G_P$ the field is swept back to the point where $G_A$ has been measured before).
  As explained above, among the formerly degenerate lowest lying states, the ones with maximum spin-projection $S_z$ are selected as ground states, as highlighted in \Tab{states}. At the first two peaks transport is mediated by $\uparrow$ -, at the last two peaks by $\downarrow$ - electrons. For parallel magnetization of the leads this breaks the mirror symmetry of the conductance by enhancing the first two peaks while suppressing the last two.
In \Fig{fig:TMRs}(c) we still neglect intrisic shifts, but assume $|h^{P}_{ext}|>|h^{A}_{ext}|\, (>k_BT$, e.g. due to stray fields) to bring also the first negative TMR mechanism, \Fig{fig:setup}(b), into play.
It overlays the effects observed in \Fig{fig:TMRs}(b). The shift in the peak positions of parallel and anti-parallel conductance peaks enforces a change of the TMR from positive to negative value from the first peak on. 
Because there is no significant change in the curves of  \Fig{fig:TMRs}(b) and \ref{fig:TMRs}(c) upon inclusion of the fourth order effects, in the remaining figures we only show second order curves. In \Fig{fig:TMRs}(d) we include the finite level shifts which arise intrinsically from our transport theory, both for $\Delta_{\rm St}=0$ 
and $\Delta_{\rm St}=0.2W$.
In the latter case we find that inclusion of excited states (here up to excitation energies of $2\epsilon_0$) \emph{can even yield negative TMR}.
Interestingly, the intrinsic gate dependent contribution distinguishes itself from all former contributions by preserving the mirror symmetry in the conductances and the TMR curve, as seen for $\Delta_{\rm St}=0$. The reason is that a given electron-hole symmetry in the tunneling results in an electron-hole symmetry for any charge fluctuation (see supplemental material). This is no longer true if $\Delta_{\rm St}\neq0$, where we observe a  behaviour similar  to that of a constant external shift but of opposite sign  ~\cite{Gaas11}. These statements are confirmed by \Fig{fig:TMRs}(e) which shows the shifts for exemplary electron-hole symmetric transitions. $\Delta_{\rm St}\neq0$ induces a gate constant positive (negative) shift, for $\uparrow$ ($\downarrow$) - mediated transitions, which breaks the electron-hole symmetry. It is nicely visible that, due to the orbital degeneracy of the CNT, the shifts \emph{differ from peak to peak}.
Finally, Fig. \ref{fig:TMRs}(f) combines the impact of the intrinsic shifts with an equal extrinsic splitting $|h^{P}_{ext}|=|h^{A}_{ext}|$.
 The intrinsic effects suffice to change the TMR curve \Fig{fig:TMRs}(b) to a shape observed in experiment~\cite{Sahoo05}, though due to our limitation to the weak coupling regime, quantitative agreement cannot be expected. An asymmetric coupling is found to affect the curve quantitatively, while the qualitative shape is retained.
 Even for the very small values of $\Gamma_s$ used here to justify lowest order perturbation theory, the difference between Figs. 2(b) and (f) reveals a marked influence of the intrinsic, tunneling induced, level shifts.

\paragraph{Summary}

We analyzed the impact of different kinds of effective Zeeman shifts in magnetically coupled multilevel quantum dots,  obtaining a characteristic gate dependence and the possibility of negative TMR. In particular, we have provided a systematic way of including the important  effects of  tunneling induced level shifts in a transport calculation.
In general, a TMR signal will be influenced by many parameters relevant to the given device. Nevertheless, following the lines of the analysis given here for a CNT, it should be possible to disentangle the importance of the various contributions.


\par

We acknowledge fruitful discussions with C. Strunk, S. Pfaller and J. Hauptmann as well as support of the DFG under the program SFB 689.


\begin{thebibliography}{8}


\bibitem{Sahoo05}
S.~Sahoo \emph{et al.},
Nat. Phys. {\bf 1}, 99 (2005).

\bibitem{Pasupathy04} A.~N. Pasupathy \emph{et al.},
Science \textbf{306}, 86 (2004).



\bibitem{Hauptmann08}
J.~R. Hauptmann, J.~Paaske, and P.~E. Lindelof, Nat. Phys. {\bf 4}, 373 (2008).

\bibitem{Hofstetter10} L. Hofstetter \emph{et al.}, Phys. Rev. Lett. \textbf{104}, 246804 (2010).

\bibitem{Gaas11}M. Gaass \emph{et al.},
arXiv:1104.5699v1

\bibitem{Cottet06b}
A.~Cottet and M.-S.~Choi,
  Phys.\ Rev.\ B {\bf 74}, 235316 (2006).

\bibitem{Cottet06}
A.~Cottet \emph{et al}.,
Semic. Sci. Tech. \textbf{21}, 78 (2006).

\bibitem{Brataas99}
A.~Brataas, Y.~V.~Nazarov, G.~E.~W.~Bauer, Phys. Rev. Lett. {\bf 84}, 2481 (1999).

\bibitem{Wetzels05}
W.~Wetzels, G.~E.~W.~Bauer and M.~Grifoni, Phys. Rev. B {\bf 72}, 020407(R) (2005).




\bibitem{Haldane78}
F.~D.~M.~Haldane, Phys. Rev. Lett. \textbf{40}, 416 (1978).

\bibitem{Martinek03} J. Martinek \emph{et al}., Phys. Rev. Lett. \textbf{91}, 127203 (2003).

\bibitem{Martinek05}J. Martinek \emph{et.~al.}, Phys. Rev. B \textbf{72}, 121302(R) 2005.




\bibitem{Koller07}
S.~Koller, L.~Mayrhofer, and M.~Grifoni, New J.
  Phys. {\bf 9}, 348 (2007).


\bibitem{Schoeller94}
H.~Schoeller and G.~Sch\"on, Phys.\ Rev.\ B {\bf 50}, 18436 (1994).

\bibitem{Mayrhofer06}
L.~Mayrhofer and M.~Grifoni, Phys.\ Rev.\ B {\bf 74}, 121403(R) (2006).








\bibitem{Koller10}
S.~Koller, M.~Leijnse, M.~R. Wegewijs and M.~Grifoni, Phys.\ Rev.\ B {\bf 82}, 235307 (2010).



\bibitem{Weymann05} I. Weymann, J. K\"onig, J. Martinek and G. Sch\"on, Phys. Rev. B \textbf{72}, 115334 (2005).
\end{thebibliography}
\end{document}


\title{SUPPLEMENTAL MATERIAL to the manuscript: Sources of negative tunneling magneto-resistance in multilevel quantum dots with ferromagnetic contacts}
\author{Sonja Koller}
\affiliation{Theoretische Physik, Universit\"at Regensburg, 93040 Regensburg, Germany}
\author{Jens Paaske}
\affiliation{The Niels Bohr Institute and Nano-Science Center, University of Copenhagen, Universitetsparken 5, DK-2100 Copenhagen {\O}, Denmark}
\author{Milena Grifoni}
\affiliation{Theoretische Physik, Universit\"at Regensburg, 93040 Regensburg, Germany}
\date{\today}
\pacs{73.63.Kv -- Quantum dots,   73.23.Hk -- Coulomb blockade, single electron tunneling, 73.63.-b Electronic transport in nanoscale materials }
\maketitle
In the main part of our Rapid Communication, we have introduced with Eq. (2) the Liouville equation for the reduced density matrix $\hrho(t) = \Tr_{\leads}\left\{\hrho_\tot(t)\right\}$ in the stationary state ($\dot{\hrho}(t)=0$),
\begin{figure}\begin{center}
\includegraphics[width=0.94\columnwidth]{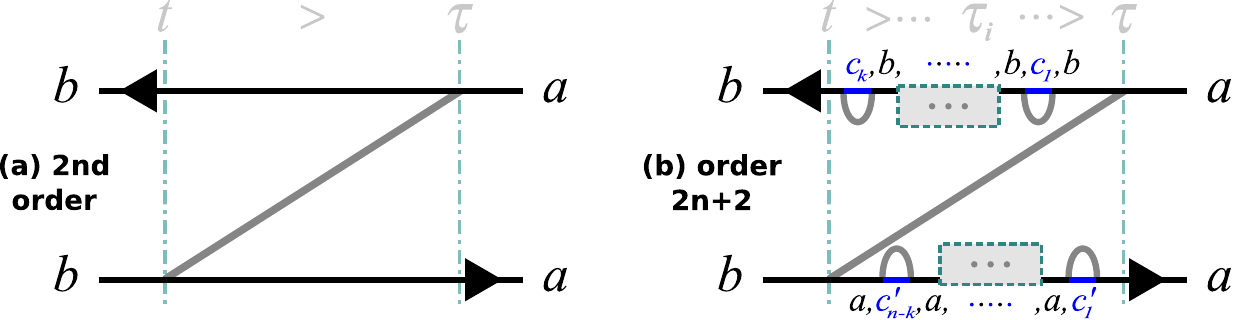}
\caption{\label{fig:ren}Structure of: (a) a sequential tunneling diagram, (b) a level renormalization diagram for a sequential tunneling event. Each "bubble" describes  a charge fluctuation. We consider  $k$ charge fluctuations (to states $c_j$) in the final state $b$ and $n-k$ charge fluctuations (to states $c'_i$) in the initial state $a$. We have indicated the time order, as well as the intermediate states on the contours.}\end{center}
\end{figure}
  \begin{equation*}
  0 = -i\sum_{aa'} \delta_{ab}\,\delta_{a'b'}\left(E_a-E_{a'}\right)
  \rho_{aa'}+\sum_{aa'}{{K}}^{aa'}_{bb'}\rho_{aa'}.
\end{equation*}
The solution to this equation describes the stationary state of the quantum-dot system in the presence of the leads.
Thereby, the main task is the determination of the kernel $K$.
A fully diagrammatic representation of $K$ has first been proposed in~\cite{Schoeller94}. For example the second order kernel
$(K^{(2)})_{bb}^{aa}$ is shown in Fig. \ref{fig:ren}(a). It describes a full tunnel event $|a\rangle\to|b\rangle$, represented by the solid line connecting upper and lower parts of the diagram contour.
 Recently,
 it has been recognized~\cite{Koller10} that certain fourth order diagrams can be related to charge fluctuations in the initial or final state during a tunnel event.
   Here we identify a specific class of these   charge fluctuation  diagrams which, summed up in all orders, yield a whole Taylor series and therewith an actual level shift.
  \Fig{fig:ren}(b)
  shows a diagram of order $2n+2$, contributing to $(K^{(2n+2)})_{bb}^{aa}$, which dresses   \Fig{fig:ren}(a) by $k$ charge fluctuations in the final state $b$ ($k$ "bubbles" on the upper part of the contour) and $n-k$ charge fluctuations in the initial state $a$ ($n-k$ "bubbles" on the lower part of the contour). We thereby look at fluctuations isolated, in the sense of separated in time, from each other: each bubble must start and end at consecutive times $\tau_i,\,\tau_{i+1}\ (1\leq i,j\leq2n-1)$. The electron transfer event is initialized at the earliest time $\tau=\tau_0$, and ending at the latest time $\tau_{2n+1}=t$. This gives $n \choose k$ possibilities for the time ordering of the bubbles among the upper and lower part of the contour. Summing all those plus their hermitian conjugates, the total contribution is, as obtained by applying diagrammatic rules~\cite{Schoeller94},
\begin{widetext}\begin{tabular}{l@{\hspace{-0.4em}}l}
\begin{minipage}{3cm}\raisebox{1.6cm}{
\includegraphics[width=3cm]{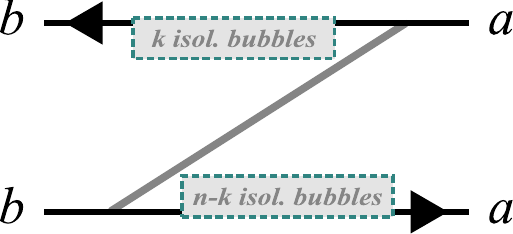}}
\end{minipage}
&
\begin{minipage}{14.6cm}
\begin{eqnarray}
\,\,\,\,+\ \ \mathrm{h.c.}\, &\sim&\lim_{\eta\to0}{n \choose k}\mathsf{Re}\, \frac{2i}{\hbar}\int\rmd\omega D_{l\sigma}(\omega) f_l^p(\omega)\left(\frac{1}{p\omega-E_{ba}+i\eta}\right)^{n+1}\left|T^{p}_{l\sigma}(b,a)\right|^2 \label{contrib}\\
&\times&\left[\prod_{j=1}^{k}\int\rmd\omega_j\frac{D_{l_j\sigma_j}(\omega_j)f_{l_j}^{p_j}(\omega_j)\left|T_{l_j\sigma_j}^{p_j}(c_j,b)\right|^2}{p_j\omega_j-E_{c_ja}
+p\omega+i\eta}\prod_{i=1}^{n-k}\int\rmd\omega_{i}\frac{D_{l_i\sigma_i}(\omega_i)f_{l_i}^{p_i}(\omega_i)\left|T_{l_i\sigma_i}^{p_i}(c'_i,a)\right|^2}{-p_i\omega_i-E_{bc'_i}+p\omega+i\eta}\right].\nonumber
\end{eqnarray}\smallskip\end{minipage}\end{tabular}
\end{widetext}
Here $c_j,\,c'_i$ serve as placeholders for the possible intermediate states; $f^p_l(\omega):=f^p(\beta\omega+\beta eV_l)$, $p=\pm$, where we defined
$f^+(\omega)= f(\omega)=1/(1+e^\omega)$, denoting the
  Fermi function, while $f^-(\omega)=f(-\omega)$.  The lead indices $l,l_i,l_j\in\{s,d\}$ are summed over.
  The values of  $p$ and $\sigma$  are set by fixing the initial and final states $a$ and $b$.
   Likewise the values of the spin indices $\sigma_j$, $\sigma_i$ and of the Fermi function labels $p_j$, $p_i$ depend on the intermediate states
   $c_j$, $c'_i$ that can be reached from $b$, $a$, respectively, and are  summed over. Here, $E_{ba}$ stands for the energy difference $E_b-E_a$ between the many body states of the quantum dot and the lead energies are integrated over. Finally, there are tunnel matrix elements characterizing transitions on the dot, $T^{-(+)}_{l\sigma}(b,a)\sim T_l(\vec{r}_l)\langle b|\psi^{(\dag)}_\sigma(\vec{r}_l) |a\rangle$ ($\vec{r}_l$ characterizes the position of lead $l$).
A multitude of terms emerges from the expression Eq. \eq{contrib} when applying the decomposition\[\lim_{\eta\to0}\int\mathrm{d}\omega\, \frac{Df^{p}(\omega)}{\omega-\mu+i\eta}=\int'\!\mathrm{d}\omega\, \frac{Df^{p}(\omega)}{\omega-\mu}-i\pi\,Df^{p}(\mu),\] ($\int^\prime$ denotes a principal part integration) for all fractions and expanding the product. We want to focus on a certain one, namely the combination where the fraction containing merely $\omega$,
\[\left(\frac{1}{p\omega-E_{ba}+i\eta}\right)^{n+1}\!\!\equiv\frac{1}{n!}\frac{\rmd}{\rmd^nE_{ba}}\,\left(\frac{1}{p\omega-E_{ba}+i\eta}\right),\]has been replaced by the delta function, and all others by their principal parts.\\
In terms of $\ f^{(n)}(\omega)=\frac{\rmd}{\rmd^n\omega}f(\omega)$ it reads:
\begin{eqnarray*}
&&\frac{2\pi}{\hbar}\frac{1}{n!} f^{(n)}(\beta E_{ba}+\beta peV_l)D_{l\sigma}\left|T_{l\sigma}^p(b,a)\right|^2\beta^n{n \choose k}\\
&&\times\mathsf{Re}\left[\prod_{j=1}^{k}\,\int^\prime\!\!\rmd\omega_j\frac{f^{+}(\omega_j)D_{l_j\sigma_j}\left|T_{l_j\sigma_j}^{p_j}(c_j,b)\right|^2}{\omega_j-\beta E_{c_jb}-p_j\beta eV_{l_j}}\right.\\
&&\left.\times\prod_{i=1}^{n-k}\,\int^\prime\!\!\rmd\omega_i\frac{f^{-}(\omega_i)D_{l_i\sigma_i}\left|T_{l_i\sigma_i}^{p_i}(c'_i,a)\right|^2}{\omega_i-\beta E_{ac'_i}+p_i\beta eV_{l_i}}\right].\\
\end{eqnarray*}
 Exhausting all possibilities of choosing the states $c_j,\,c'_i$ as well as setting $k=1\cdots n$, it is clear that one
generates all kinds of terms appearing in an expansion of the power $n$ of a quantity $h_{int}^{ba}$ defined as
\begin{multline}h_{int}^{ba}=\sum_{l}\left(\sum_{c}\Bigl|T_{l\sigma}^{p}(c,b)\Bigr|^2
\int^\prime \!\!\rmd\omega \frac{f^{+}(\omega)D_{l\sigma}(\omega)}{\omega-\beta E_{cb}-p\beta eV_{l}}\right.\\ \left.
+\sum_{c'}\Bigl.\Bigl|T_{l\sigma}^{p}(c',a)\Bigr|^2\int^\prime \!\!\rmd\omega \frac{f^{-}(\omega)D_{l\sigma}(\omega)}{\omega-\beta E_{ac'}+p\beta eV_{l}}\right),\label{splittingtot}
\end{multline}
where $c$ and $c'$ run over all states connected to $|b\rangle$, respectively $|a\rangle$ via a charge fluctuation.
In total, we obtain the $n$th term $\frac{2\pi}{\hbar}\frac{1}{n!} f^{(n)}(\beta(E_{ba}+peV_l))\left|T_{l\sigma}^p(b,a)\right|^2D_{l\sigma}(\beta h_{int}^{ba})^n$
in the Taylor expansion of a Fermi function $\frac{2\pi}{\hbar}f(\beta(E_{ba}+eV_l+h_{int}^{ba}))\left|T_{l\sigma}^p(b,a)\right|^2D_{l\sigma}$. So, effectively, the considered contribution of the initial and final state charge fluctuations renormalizes any energy difference $E_{ba}$ to $E_{ba}+h_{int}^{ba}$. From comparison with \Eq{contrib}, we can extract a renormalization to a manybody energy $E_b$ by looking at all the possible fluctuations on the upper contour:
\begin{equation}
\delta E_b=\sum_{l}\sum_{c}\Bigl|T_{l\sigma}^{p}(c,b)\Bigr|^2
\int^\prime \!\!\rmd\omega \frac{f^{p}_l(\omega)D_{l\sigma}(\omega)}{p\omega- E_{cb}},
\label{splittingSIAM}
\end{equation}
where the values of $p$ and $\sigma$ are fixed by the diagrammatic rules \cite{Schoeller94} once the state $c$ is assigned.
For the case of the single impurity Anderson model this expression reduces to the result for the level shift  obtained in perturbation theory \cite{Haldane78,Martinek03}. Hence Eqs. (2), (\ref{splittingSIAM}) provide a diagrammatic interpretation of
these results.
For a multilevel quantum dot, the summation over the intermediate virtual states $c$ implies that the amplitude of the level shifts is
expected to vary from a resonance peak to the other, as also shown in Fig. 2(e) of the main part of the manuscript. Notice also that for
the CNT used in our calculations, the summation over the virtual states also include states with bosonic excitations as described by the
bosonic Hamiltonian $\hat H_B$ in Eq. (2) of the main part of manuscript.

Having identified the underlying diagrams, it is now clear how to include the effects of tunneling induced level shifts quite generally in
a diagrammatic evaluation of the  kernel $K$. As temperature goes to zero, these diagrams diverge logarithmically at the charge-degeneracy points, and hence give a parametric selection of these specific diagrams. In general,  diagrams with multiple single charge excitations are  expected to give the dominant contribution to the charge fluctuations in the presence of interactions. This infinite class of diagrams is seen to correspond to a perturbative renormalization of the dot spectrum, an effect which is indeed expected on physical grounds. In conventional Feynman diagrammatics, these terms arise from summing up the 2nd order dot electron self-energy in the Dyson equation.

In order to better analyze the contribution of majority and minority spins we consider the case of a flat band including a Stoner splitting $\Delta_{\rm St}$. The density of states assumes the form \cite{Martinek05,Sindel07}
\begin{equation}
D_{l\sigma}(\omega)=D_{l\sigma}\theta(\omega+W+\sigma\Delta_{\rm St}/2)\theta(W-\sigma\Delta_{\rm St}/2-\omega),\label{DOS}
\end{equation}
where $D_{l\sigma}=D_l(1+\sigma P_l)$ is the density of states at the Fermi level, and $P_l$  the polarization of contact $l$.\\
Notice that in order to retain holomorphic functions in \Eq{splittingtot}, one can make use of the Fermi function representation of the step functions, $\theta(\pm x)=\lim_{\beta\to \infty}f^{\mp}(\beta x)$. Using then the residue calculus \cite{Seb}, one obtains
 \begin{multline}h^{ba}_{int}=\sum_{l}\left(\sum_{c}D_{l\sigma}\Bigl|T_{l\sigma}^{p}(c,b)\Bigr|^2\digPsi_{\sigma,-}\Bigl(\beta E_{cb}+p\beta eV_{l}\Bigr)\right.\\ \left.
 -\sum_{c'}D_{l\sigma}  \Bigl|T_{l\sigma}^{p}(c',a)\Bigr|^2\digPsi_{\sigma,+}\Bigl(\beta E_{ac'}-p\beta eV_{l}\Bigr)\right), \label{h_tot_ex}\end{multline}
where the upper/lower boundary $\pm W-\sigma\Delta_{\rm St}/2$ of the integration enters via
\begin{multline*}\tilde{\Psi}_{\sigma,\pm}^{(0)}(x)\equiv\Re\left[\Psi^{(0)}\left(0.5+\frac{ix}{2\pi}\right)\right.\\\left.-\Psi^{(0)}\left(0.5+i\frac{x-\beta(\pm W - \sigma\Delta_{\rm St}/2)}{2\pi}\right)\right],\end{multline*}
where $\Psi^(0)$ is the digamma function. An approximation of this result is obtained elegantly when using \Eq{DOS} to split the integration range of the
 integrals in Eq. (\ref{splittingtot}) as
 \begin{multline}
 \int_{-W-\Delta_{\rm St}/2}^{W-\Delta_{\rm St}/2}d\omega = \int_{-W-\Delta_{\rm St}/2}^{-W+\Delta_{\rm St}/2}d\omega + \int_{-W+\Delta_{\rm St}/2}^{W-\Delta_{\rm St}/2}d\omega \,,\nonumber\\
 \int_{-W+\Delta_{\rm St}/2}^{W+\Delta_{\rm St}/2}d\omega = \int_{-W+\Delta_{\rm St}/2}^{W-\Delta_{\rm St}/2}d\omega + \int_{W-\Delta_{\rm St}/2}^{W+\Delta_{\rm St}/2}d\omega \,,
 \end{multline}
 and noting that in the region at the top/bottom of the band is $f(\omega)=0/1$. Here  we get the result
 \begin{multline}h^{ba}_{int}=\sum_{l}\left(\sum_{c}D_{l\sigma}\Bigl|T_{l\sigma}^{p}(c,b)\Bigr|^2\digPsi_{\downarrow,+}\Bigl(\beta E_{cb}+p\beta eV_{l}\Bigr)\right.\\ \left.
 -\sum_{c'}D_{l\sigma}  \Bigl|T_{l\sigma}^{p}(c',a)\Bigr|^2\digPsi_{\uparrow,-}\Bigl(\beta E_{ac'}-p\beta eV_{l}\Bigr)\right)\\
 +\sum_{l,c} D_{l\uparrow} \Bigl|T_{l\uparrow}^{p}(c,b)\Bigr|^2 \ln\left(\frac{W_\uparrow+ E_{cb}+p eV_l}{W_\downarrow+ E_{cb}+p  eV_l}\right)
\\
+\sum_{l,c'}D_{l\downarrow} \Bigl|T_{l\downarrow}^{p}(c',a)\Bigr|^2 \ln\left(\frac{W_\downarrow- E_{ac'}+p eV_l}{W_\uparrow- E_{ac'}+p  eV_l}\right), \label{h_tot}\end{multline}
where $W_\sigma=W-\sigma\Delta_{\rm St}/2$ is the finite bandwidth for the integration over the lead energies. The result is an approximation of \Eq{h_tot_ex}, as can be seen when applying $\ln(A/B)\approx\Re\Psi(0.5+iA)-\Re\Psi(0.5+iB)$ in \Eq{h_tot}.

Obviously, the last two terms in Eqs. (\ref{h_tot_ex}), (\ref{h_tot}) vanish if $\Delta_{\rm St}=0$. Moreover, for $E_{cb}$, $E_{ac'} \ll W_\sigma $ they are very weakly dependent on the gate voltage and polarization, but dependent on the saturation magnetization via the Stoner splitting. Following \cite{Gaas11}, with tunneling coupling $\Gamma\simeq 1 $meV \cite{note}, this contribution can yield for PdNi contacts the gate-independent part of the level shift to be in the order of few hundreds of $\mu$eV. This is in the same order as the constant shift used to fit the Kondo data in \cite{Gaas11,Hauptmann08} but also the TMR data \cite{Sahoo05} in \cite{Cottet06b}.
In contrast, the first two terms  depend strongly on gate voltage, interaction and polarization.
They correspond to the contribution to the level shift first proposed in \cite{Martinek03}. 
This allow us to recast the expression for $h^{ba}_{int}$ in the form
\begin{equation}
h^{ba}_{int}= h^{ba}_{0}+\Delta h^{ba}_{},
\label{h0+hcost}
\end{equation}
where $\Delta h^{ab}_{}$ is the contribution due to $\Delta_{\rm St}$.

If the states $a,b,c,c'$ have electron-hole symmetry partners $a_\star,b_\star,c_\star,c'_\star$, there exists the mirror transition $b_\star\to a_\star$, with $a_\star$ now being the final, $b_\star$ the initial state. Thus the contributions from the gate dependent part of the charge fluctuations to $c_\star$ is negative, to $c'_\star$ positive, such that $h_{0}^{b_\star a_\star}=-h_{0}^{ba}$. This preserves the mirror symmetry of all curves when $\Delta_{\rm St}=0$.

Importantly, this concept generalizes to higher orders:  virtual charge fluctuations during inelastic cotunneling (appearing first in sixth order) give a renormalization of the inelastic cotunneling threshold, as already experimentally observed in \cite{Holm08}.
In the present work the constant part of the level shift has been implemented in the numerical calculation of the 4th order contribution to the TMR shown in Figs. 2(b) and 2(c) of the main part of this Rapid Communication.  However, in the considered weak coupling regime  there is no significant change in the curves of Fig. 2(b) and 2(c) upon inclusion of the fourth order effects. Because the numerical implementation of the gate dependent contribution $h^{ba}_0$ implies a
summation over a large number of states for a CNT quantum dot and no significative changes are expected,
  in the remaining figures Fig. 2(d) - (f) only the results of a second order calculation with  all level shifts fully included are shown.